\documentclass[12pt,preprint]{aastex}
\usepackage{amsmath,amssymb,natbib,graphicx}
\usepackage{latexsym}
\usepackage{graphicx}
\usepackage{threeparttablex}
\usepackage{
lscape, longtable}
\usepackage{geometry}
\usepackage[T1]{fontenc}
\usepackage{multirow}
\DeclareGraphicsExtensions{.png,.jpg}
\shorttitle{Disk Detective}
\shortauthors{Kuchner, et al.}
\begin{document}

\slugcomment{{\sc Accepted for publication in {\it ApJ}:} July 17, 2016} 

\title{Disk Detective: Discovery of New Circumstellar Disk Candidates through Citizen Science}

\author{Marc J. Kuchner}
\affil{NASA Goddard Space Flight Center 
\\ Exoplanets and Stellar Astrophysics Laboratory, Code 667
\\ Greenbelt, MD 21230}
\email{Marc.Kuchner@nasa.gov}

\author{Steven M. Silverberg }
\affil{Homer L. Dodge Department of Physics and Astronomy
\\ The University of Oklahoma
\\ 440 W. Brooks St.
\\ Norman, OK 73019}
\email{silverberg@ou.edu}

\author{Alissa S. Bans}
\affil{Adler Planetarium 
\\ 1300 S Lake Shore Dr
\\ Chicago, IL 60605}
\email{abans@uchicago.edu}

\author{Shambo Bhattacharjee}
\affil{School of Statistics,
\\ University of Leeds
\\ Leeds LS2 9JT, England}
\email{shambo.bhattacharjee@gmail.com}

\author{Scott J. Kenyon}
\affil{Smithsonian Astrophysical Observatory
\\ 60 Garden Street
\\ Cambridge, MA 02138 USA}
\email{skenyon@cfa.harvard.edu}

\author{John H. Debes}
\affil{Space Telescope Science Institute 
\\ 3700 San Martin Dr.
\\ Baltimore, MD 21218}
\email{debes@stsci.edu}

\author{Thayne Currie}
\affil{National Astronomical Observatory of Japan
\\ 650 N A'ohokhu Place
\\ Hilo, HI 96720}
\email{thayne.currie@gmail.com}

\author{Luciano Garcia}
\affil{Observatorio Astron\'omico de C\'ordoba
\\Universidad Nacional de C\'ordoba
\\Laprida 854, X5000BGR, C\'ordoba, Argentina}
\email{lucianog@oac.uncor.edu}

\author{Dawoon Jung}
\affil{Korea Aerospace Research Institute
\\ Lunar Exploration Program Office
\\ 169-84 Gwahak-ro, Yuseong-gu, Daejeon 34133 Korea}
\email{dwjung@kari.re.kr}

\author{Chris Lintott}
\affil{Denys Wilkinson Building
\\ Keble Road
\\ Oxford, OX1 3RH}
\email{cjl@astro.ox.ac.uk}

\author{Michael McElwain}
\affil{NASA Goddard Space Flight Center 
\\ Exoplanets and Stellar Astrophysics Laboratory, Code 667
\\ Greenbelt, MD 21230}
\email{michael.w.mcelwain@nasa.gov}

\author{Deborah L. Padgett}
\affil{NASA Goddard Space Flight Center 
\\ Exoplanets and Stellar Astrophysics Laboratory, Code 667
\\ Greenbelt, MD 21230}
\email{deborah.l.padgett@nasa.gov}

\author{Luisa M. Rebull}
\affil{Infrared Processing and Analaysis Center 
\\ Caltech M/S 314-6          
\\ 1200 E. California Blvd.   
\\ Pasadena, CA 91125}
\email{rebull@ipac.caltech.edu}

\author{John P. Wisniewski}
\affil{Homer L. Dodge Department of Physics and Astronomy
\\ The University of Oklahoma
\\ 440 W. Brooks St.
\\ Norman, OK 73019}
\email{wisniewski@ou.edu}

\author{Erika Nesvold}
\affil{Department of Physics, University of Maryland Baltimore County 
\\ 1000 Hilltop Circle
\\ Baltimore, MD 21250}
\email{Erika.Nesvold@umbc.edu}

\author{Kevin Schawinski}
\affil{ETH Z{\"u}rich Institute for Astronomy
\\ Wolfgang-Pauli-Strasse 27
\\ Building HIT, Floor J
\\ CH-8093 Z{\"u}rich, Switzerland}
\email{kevin.schawinski@phys.ethz.ch}

\author{Michelle L. Thaller}
\affil{NASA Headquarters
\\ Science Mission Directorate
\\ 300 E St SW
\\ Washington, D.C. 20546}
\email{michelle.thaller@nasa.gov}

\author{Carol A. Grady}
\affil{NASA Goddard Space Flight Center 
\\ Exoplanets and Stellar Astrophysics Laboratory, Code 667
\\ Greenbelt, MD 21230}
\email{carol.a.grady@nasa.gov}


\author{Joseph Biggs}
\affil{Disk Detective}

\author{Milton Bosch}
\affil{Disk Detective}

\author{Tadeas Cernohous}
\affil{Disk Detective}

\author{Hugo A. Durantini Luca}
\affil{Disk Detective}

\author{Michiharu Hyogo}
\affil{Disk Detective}

\author{Lily Lau Wan Wah}
\affil{Disk Detective}

\author{Art Piipuu}
\affil{Disk Detective}

\author{Fernanda Pineiro}
\affil{Disk Detective}

\begin{abstract}
The Disk Detective citizen science project aims to find
new stars with 22 $\mu$m excess emission from circumstellar dust using
data from NASA's WISE mission. 
Initial cuts on the AllWISE catalog provide an input catalog of 277,686 sources.
Volunteers then view images of each source online in 10 different bands to
identify false-positives (galaxies, background stars, interstellar matter, image artifacts, etc.).
Sources that survive this online
vetting are followed up with spectroscopy on the FLWO Tillinghast telescope.
This approach should allow us to unleash the full potential of WISE for finding new debris
disks and protoplanetary disks.  We announce a first list of 37 new disk candidates discovered
by the project, and we describe our vetting and follow-up process.
One of these systems appears to contain the first debris disk discovered
around a star with a white dwarf companion: HD 74389.
We also report four newly discovered classical
Be stars (HD 6612, HD 7406, HD 164137, and HD 218546) and a new
detection of 22 $\mu$m excess around a previously known debris disk host star HD 22128.
\end{abstract}

\section{Introduction}
\label{sec:introduction}

All-sky mid-infrared surveys have revolutionized the science of planet formation by discovering populations of young stars and main sequence stars with excess infrared radiation indicating the presence of dusty circumstellar disks. These disks, which include gas-rich protoplanetary disks around Young Stellar Objects (YSOs) and dusty debris disks around main sequence stars, serve as the signposts of planet formation \citep[e.g.,][]{2002ApJ...577L..35K}.  They inform us about the timescales and the environment of planet formation \citep[e.g.,][]{2005ApJ...620.1010R, 2007ApJ...671.1784H, 2015ApJ...808..167J}, and the present day locations and dynamics of planets \citep[e.g.,][]{2010ApJ...718L..87T, 2012ApJ...748L..22M, 2013ApJ...766L...1Q,  2015ApJ...798...83N, 2015ApJ...807L...7C}.  

The IRAS all sky survey discovered the first extrasolar debris disks \citep{1984ApJ...278L..23A} and provided a large sample of debris disks \citep[e.g.,][]{2007ApJ...660.1556R}.  After IRAS, AKARI surveyed the whole sky at 9 and 18 $\mu$m with $\sim 7$ times better sensitivity than IRAS, finding many more new disks \citep{2010A&A...514A...1I}.   Some disk discoveries have come from pointed studies, like the Spitzer Formation and Evolution of Planetary Systems (FEPS) survey \citep{2009ApJS..181..197C}. But many of the best-studied, most informative disks (like TW Hydra, Fomalhaut, etc.) are relatively isolated on the sky, requiring an all-sky survey to find them.  

NASA's Wide-field Infrared Survey Explorer (WISE) is the most recent and sensitive all-sky mid-infrared survey \citep{2010AJ....140.1868W}, with a further factor of $\sim 80$ gain in sensitivity over AKARI in the mid-IR.  Using a 16-inch mirror in a Sun-synchronous polar orbit, WISE scanned the sky at 3.4 $\mu$m, 4.6 $\mu$m, 12 $\mu$m, and 22 $\mu$m  (bands W1, W2, W3, and W4 respectively).  The WISE cryogenic mission, launched in 2009, lasted a little over 10 months and was followed by the first post-cyrogenic mission, NEOWISE. The AllWISE catalog\footnote{Available at http://irsa.ipac.caltech.edu/Missions/wise.html} combines data from both phases, making it the most comprehensive mid-infrared multi-epoch view of the sky available today. 

Previous infrared surveys for debris disks have provided target lists for exoplanet searches via direct imaging \citep{2008ApJ...672.1196A, 2013ApJ...773...73J, 2013ApJ...773..179W, 2015ApJ...800....5M}.  Debris disks found with WISE should provide crucial targets for upcoming generations of exoplanet searches.  WISE could detect debris disks around main sequence A stars to a distance of 300 pc and protoplanetary disks around T Tauri stars to 1 kpc.

Indeed, many teams have used the WISE data to find new debris disks, searching a vast catalog of $>747$ million WISE sources. \citet{2012MNRAS.427..343M} cross-correlated the WISE source list with the Hipparcos catalog, finding over 86,000 stars with suspected infrared excesses.  \citet{2013MNRAS.433.2334K}, \citet{2013ApJS..208...29W} and \citet{2014ApJS..212...10P} performed more careful searches for debris disks in the WISE source list using the Hipparcos catalog and found 6, 70 and 108 new debris disk candidates, respectively.  Other specific surveys for debris disks have focused on stars with ages determined from chromospheric activity \citet{2014ApJ...780..154V}, white dwarfs \citep{2011ApJ...729....4D, 2012ApJ...759...37D}, M dwarfs \citep{2012A&A...548A.105A, 2013AAS...22115813O}, G-K dwarfs \citep{2014MNRAS.437..391C}, Kepler candidate exoplanet systems \citep{2012ApJ...752...53L, 2012A&A...541A..38R, 2012MNRAS.426...91K} and other exoplanet catalogs \citep{2012ApJ...757....7M}. 

Likewise, the WISE data on young clusters and star-forming regions have attracted much attention. \citet{2011ApJS..196....4R, 2014ApJ...784..126E, 2014AJ....147..133L} scoured the Taurus-Auriga Region. \citet{2012ApJ...744..130K} searched 11 outer Galaxy massive star-forming regions and three open clusters. Other studies have examined smaller regions, like the Western Circinus molecular cloud \citep{2011ApJ...733L...2L}, the young open cluster IC 1805 \citep{2013A&A...554A...3S} the \ion{H}{2} region S155 \citep{2014RAA....14.1269H}, the Sco-Cen and $\eta$ Cha associations \citep{2012MNRAS.421L..97R, 2012ApJ...758...31L}, nearby moving groups of young stars \citep{2012ApJ...751..114S} and $\lambda$ and $\sigma$ Orionis \citep{2015AJ....150..100K}. Still others have attempted to take in the whole sky, using color cuts \citep{2013Ap&SS.344..175M} or cross correlating with IRAS \citep{2014ApJ...784..111L}. Many of these searches were based on preliminary data releases with less sensitivity than the AllWISE release, but they have already uncovered thousands of candidate Class I, II and III YSOs and transitional disks, helping fill in our picture of the timing and progression of star formation.  

Unfortunately, because of its limited spatial resolution (12 arcsec at 22 $\mu$m) contamination and confusion limit every search for disks with WISE \citep[e.g.][]{2012MNRAS.426...91K}. Contamination sources include unresolved companion stars and other stars nearby on the sky, background galaxies, Galactic cirrus, and even asteroids and airplanes. For this reason, most recent searches include visual inspection of the WISE images \citep[see e.g.,][]{2011ApJ...729....4D, 2013ApJS..208...29W, 2014MNRAS.437..391C, 2014ApJS..212...10P}. Computer cuts alone can provide a first stage of vetting, but they generate catalogs riddled with false positives \citep{2012MNRAS.426...91K}; .  Color cuts and source quality flags can help \citep[e.g.,][]{2012ApJ...744..130K, 2013Ap&SS.344..175M, 2014MNRAS.440.3430D}, but the color loci of disk candidates overlaps with the color loci of blended background galaxies \citep{2012ApJ...744..130K} and peaks in the Galactic dust emission.  \citet{2012MNRAS.426...91K} used the IRAS 100 $\mu$m level to discard many false positive disk candidates contaminated with Galactic dust emission, but using this method prohibits searching many interesting star-forming regions. 

Because of these challenges, many disks remain to be found with WISE data, even after all the efforts described above. The largest published study of debris disks \citep{2013ApJS..208...29W} and the still larger WISE science team disk study (Padgett et al. in prep.) are based on the Hipparcos and Tycho catalogs. These catalogs are magnitude limited in V band, so they omit a vast population of redder, late type stars\footnote{The initial candidates listed in this paper are also all in Hipparcos, but our full dataset does not cross-correlate with any stellar catalogues.}. 

Moreover, a vast solid angle in young clusters and star-forming regions remains to be properly searched with WISE---each candidate examined by eye and followed up with spectroscopy and higher resolution imaging.  When \citet{2013Ap&SS.344..175M} ran the all-sky data through a novel color filter to search for YSO candidates (without vetting the candidates by eye), he found a total of $\sim 10,000$ objects of interest; the WISE studies of young clusters and star-forming regions described above (which mostly included visual vetting) yielded a total of $\sim 4000$ disk candidates. The difference between these numbers provides a minimal measure of what remains for us to study with WISE: $\gtrsim 6000$ objects with colors consistent with YSOs that have not yet been visually inspected.

Here we describe a new project to scour the WISE data for new debris disks and YSOs.  The Disk Detective citizen science/crowdsourcing project classifies WISE sources via a website, diskdetective.org, where volunteers examine images from WISE, the Two Micron All Sky Survey (2MASS), the Digitized Sky Survey (DSS) and when available, the Sloan Digital Sky Survey (SDSS), to check them for false positives. This approach should allow us to unleash the full potential of WISE for finding new  disks, probing the cooler stars and isolated objects missed by previous debris disk searches, a catalog 8 times the size of the large \citet{2013ApJS..208...29W} survey. We describe the online vetting process in Section \ref{sec:approach}, our small-telescope follow-up program in Section \ref{sec:followup} and we present our first list of 37 disk candidates in Section \ref{sec:candidates}.

\section{Citizen Science Approach}
\label{sec:approach}

\begin{figure*}
\begin{center}
\includegraphics[ totalheight=.6\textheight, angle=90]{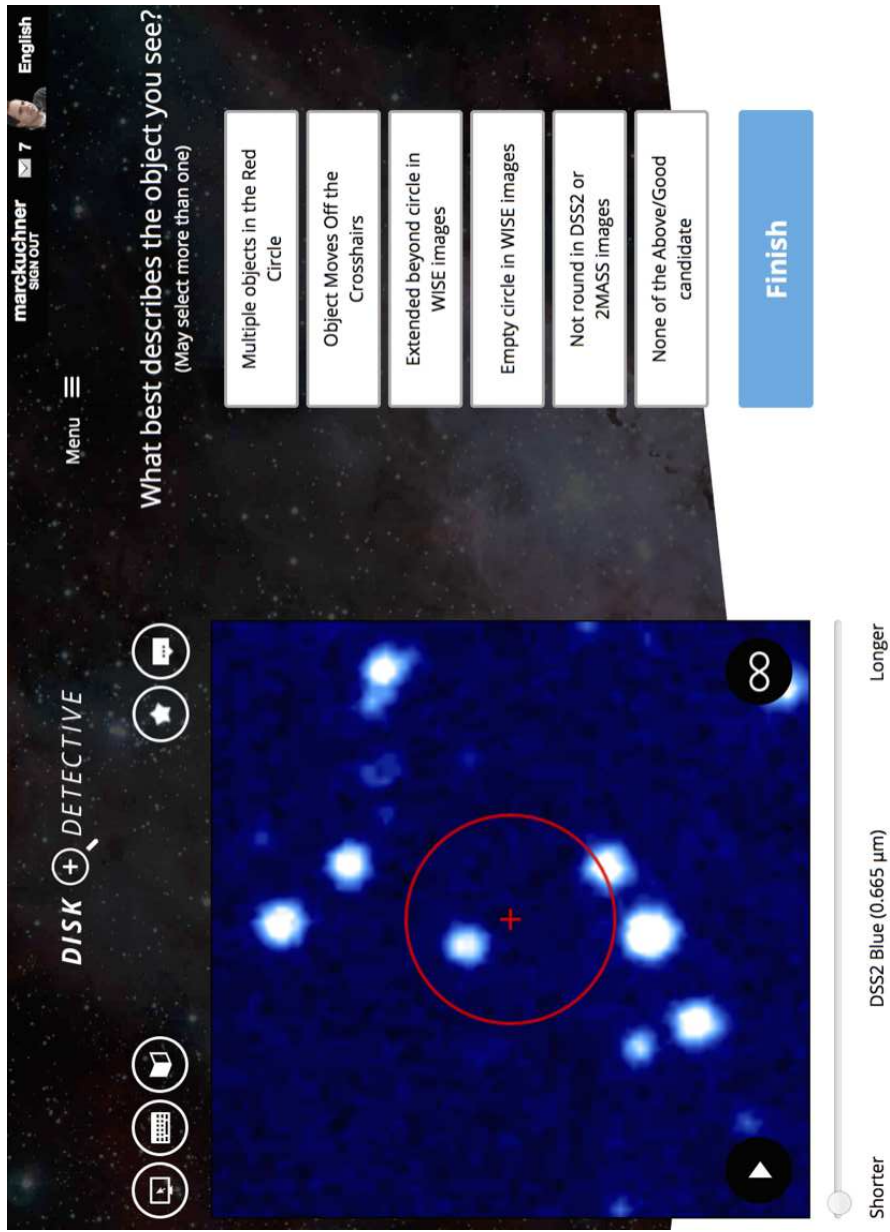}
\caption{Screen shot from DiskDetective.org showing one frame from the flipbook on the left and the classification buttons on the right. This example clearly has multiple objects visible in the DSS2 Blue band that fall within the WISE 4 beam (the red circle); the website makes it easy for participants to discard this potential false positive by clicking on the appropriate button.  \label{Fig_screenshot}}
\end{center}
\end{figure*}

Disk Detective  is  a  new  addition  to  the  successful  Zooniverse network of Citizen Science Alliance projects \citep{2008MNRAS.389.1179L}. Visitors to the site (``users'') view ``flipbooks'' showing several images of the same source at different wavelengths.  Figure~\ref{Fig_screenshot} shows a sample screenshot from DiskDetective.org illustrating one frame in a flipbook.

After they view the flipbooks, users answer a question, ``What best describes the object you see?'', by clicking on one or more of six buttons.   The site then records the user's choice(s) for interpretation by the Disk Detective science team and offers the user another source to classify.  This approach is borrowed from another Zooniverse Project, Snapshot Serengeti (http://www.snapshotserengeti.org). Snapshot Serengeti shows users flipbooks of wildlife photographs, asking users to identify animal and bird species that appear in the images, taking advantage of the human eye's ability to spot moving objects camouflaged by noise \citep{2015Swanson}.  At Disk Detective, instead of identifying animals in the grasses of the Serengeti, users identify clean point sources in a forest of astrophysical and instrumental contaminants: galaxies, ISM, artifacts, etc. 

The flipbooks in Disk Detective generally consist of ten images of each source: images in four WISE bands, three bands from the 2-Micron All Sky Survey \citep[2MASS]{2006AJ....131.1163S} and three bands from the Digital Sky Survey \citep[DSS]{1998wfsc.conf...89D}. When possible, we use images from the SDSS seventh data release instead of DSS; these SDSS data cover about 1/4 of the sky \citep[see][]{2006AJ....131.2332G, 2009ApJS..182..543A}. The images are independently scaled using IDL color table 1 (BLUE/WHITE), matching the color scale to the full range of the data. Hence, the flipbook approach works better than attempting to show all these bands in a single multi-color image, which would tend to be dominated by one or two bands.  Overlaid on every flipbook is a circle with a radius of 10.5 arcseconds, roughly the area we must ensure is free of contamination before we can trust the photometry.  (The circle's radius was chosen conservatively; the nominal resolution of WISE at 22 $\mu$m is 12 arcseconds FWHM.) Also, overlaid on each image is a cross marking the center of the image at the WISE 3.4 $\mu$m band (W1) and indicating the expected uncertainty in position for an uncontaminated point source. The disk candidates are unresolved by WISE, so our search image is a tightly concentrated red object, possibly with diffraction spikes, that does not shift position from band to band.

Users view the flipbook by clicking on an arrow-shaped ``play'' button on the lower left of the screen, shown in Figure~\ref{Fig_screenshot}. They also have the option to scroll through the flipbook frame by frame using the scrub bar beneath it.  The frames are each labeled with the wavelength and the name of the survey that yielded the image, e.g., ``2MASS K (2.16 $\mu$m)''.  After a user has viewed the whole flipbook, he or she can then choose from among six classification buttons to click on, labeled ``Multiple objects in the Red Circle'', ``Object Moves off the Crosshairs'', ``Extended beyond circle in WISE Images'' ``Empty Circle in WISE images'', ``Not Round in DSS2 or 2MASS images'' and ``None of the Above/Good Candidate''.  With the exception of the ``None of the Above'' option, the user can choose more than one description per flipbook.  After at least one of these classification buttons is chosen, a button labeled ``Finish'' becomes active; clicking this button records the user's choices and causes the next flipbook to appear.

Since its launch, Disk Detective has attracted a vast user community. Roughly 1.5 million classifications have been performed so far, by roughly 28,000 volunteers. Roughly half of the classifications come from an enthusiastic group of ``superusers''. Fifteen superusers have already classified $>10,000$ sources each; seven have classified $> 30,000$. The superusers started their own email discussion group via Google groups to work together on Disk Detective issues. They now help train other users and form a crucial extension of the Disk Detective science team (see below).

Communication with Disk Detective users is aided by the new Zooniverse translation crowdsourcing tool. Using this online tool, volunteers have translated the site into Spanish, French, Russian, German, Hungarian, Polish, Bahasa, Romanian, Portuguese, Japanese, and Chinese (traditional and simplified characters); the translated sites are accessible via a link in the upper right corner of site. 

The DiskDetective.org site is tied into the ``TALK'' social network common to Zooniverse sites.  TALK has a page for each subject on Disk Detective that provides a simple Spectral Energy Distribution (SED) for the object composed of the 2MASS and AllWISE photometry and also a link to the SIMBAD page on the source if one exists.  On TALK, users can create and maintain collections of their favorite subjects by clicking on a button labeled ``collect''.

\subsection{Pre-Selection of WISE Sources}
\label{sec:approach2}

To choose sources to upload to the website, we performed some initial computer-based cuts on the WISE data, informed by published debris disk searches (see above) but not limited by the Hipparcos or Tycho catalogs, for example. We utilized signal-to-noise cuts (\verb"w4snr, w1sigmpro w4rchi2 and w4sigmpro") and some of the AllWISE
catalog flags (\verb"cc_flags, xscprox, na, nb, n_2mass, and ext_flg") to remove sources that were noisy or close to known extended sources. Though many searches have used sophisticated color cuts to focus on particular kinds of disks \citep[e.g.,][]{2012ApJ...744..130K}, we kept our color cuts minimal to cast as broad a net as possible. To preselect sources with infrared excesses, we merely removed all sources with [W4]  $>$ [W1]-0.25. ([W1] is magnitude in the WISE 3.4 $\mu$m band, and [W4] is magnitude in the WISE 22 $\mu$m band.) This criterion corresponds to a 26\% excess over a Rayleigh-Jeans slope between those two bands. We also required that a WISE 4 excess significant at the 5-$\sigma$ level. Table~\ref{tab:cuts} summarizes all these initial cuts.

\begin{table*}
	\begin{tabular} {l l}
	Criterion	& Meaning	 \\
	\tableline
	w1mpro $>$ 3.5   & WISE 1 magnitude $> 3.5$ 			 \\
	w4mpro $<$ (w1mpro - 0.25)  & WISE 4 excess of 0.25 magnitudes over W1  \\
       w1mpro $>$ 5*sqrt(w1sigmpro*w1sigmpro  &   The WISE 4 excess is significant at the 5-$\sigma$ level. \\
         ~~+ w4sigmpro*w4sigmpro) + w4mpro  &   \\
	w4sigmpro is not null and w4rchi2 $<$ 1.3  & WISE 4 profile-fitting yielded a fit with $\chi^2 < 1.3$ \\
         w4snr $>=$ 10  & WISE 1 profile-fit signal-to-noise ratio $>$ 10  \\
         w4nm $>$ 5  & Source detected at WISE 4 in at least 5 individual \\ 
                               & ~~~8.8 second exposures with SNR$>3$  \\ 
         na = 0 and nb = 1 & The profile-fitting did not require active deblending. \\
         n\_2mass = 1  & One and only one 2MASS PSC entries found \\ 
                                 & ~~~within a 3" radius of the WISE 1 source position.  \\ 
         cc\_flags[1] not matches '[DHOP]'    & No diffraction spike, persistence, halo or optical ghost \\
         cc\_flags[4] not matches '[DHOP]'    & ~~issues at WISE 1 or WISE 4. \\
        xscprox is null or xscprox $>$ 30  & No 2MASS XSC source $< 30''$ from the WISE source. \\
        ext\_flg = 0    & Photometry not contaminated by known\\
                              &~~~2MASS extended sources. \\
	\end{tabular}
\caption{Initial cuts on the AllWISE source catalog.}
\label{tab:cuts}
\end{table*}

We launched the site on January 28, 2014 with a first batch of subjects covering only the Galactic latitudes +30 to +40,  +50 to +90, and -40 to -90, with no additional magnitude limit.  At first, about 20\% of the initial upload was made available on the site.  We soon realized that most of the volunteer effort was being spent classifying faint, extragalactic sources, so we decided to impose a magnitude limit on the search.  We chose a criterion of $J < 14.5$ because the subjects brighter than this magnitude were clearly concentrated in the Galactic plane, while fainter subjects appeared to be isotropically distributed on the sky.  We uploaded a second batch of 272,022 subjects on May 30, 2014.  This second batch covered the rest of the sky, but was limited to $J < 14.5$.  We also deactivated all the previously uploaded and active $J > 14.5$ subjects on this date, so presently the $J > 14.5$ subjects currently make up a small subset of the data for which we have classifications.

\subsection{Classification Data}

We examine the classification data weekly to chart our progress. As of August 25, 2015, the selection of classification buttons had the distribution shown in Table~\ref{tab:buttons}.  Clearly, the dominant false positive rejected by the classification process is ``Multiple objects in the Red Circle''. 

Note that the ``Empty Circle in WISE images'' button exists mainly to allow users a reasonable response option if, for example, there were a network glitch. None of the data actually had empty circles in all the WISE images, though some subjects did have empty red circles in a single band, and some users chose this ``Empty Circle in WISE images'' classification for such subjects.  The rarity of this situation is reflected in the very low (0.6\%) classification rate for this button.

\begin{table*}
	\begin{tabular} {l c c}
				&  \multicolumn{2}{c}{Percentage of Times Selected} \\
        Classification Button  			                        & J  < 14.5  & J > 14.5  \\
	\tableline
	Multiple objects in the Red Circle 		& 40.5 & 6.5 \\
	Object Moves Off the Crosshairs 				& 6.5 & 6.9\\
	Extended beyond circle in WISE Images 				& 23.4 & 15.9 \\
	Empty circle in WISE images 	                     & 0.6 & 0.4 \\
	Not round in DSS2 or 2MASS images 		& 9.0 & 37.6 \\
	None of the Above/Good Candidate 				& 20.0 & 33.0  \\
	\end{tabular}
\caption{Classification buttons, and the distribution of raw classifications as of August 25, 2015. The distribution is broken down by J magnitude of the sources.  The bright objects were more frequently classified as ``Multiple objects in the Red Circle''; the fainter ones as ``Not round in DSS2 or 2MASS images''. We deactivated the J > 14.5 objects on May 30, 2014.}
 \label{tab:buttons}
\end{table*}

So far, we have investigated two basic algorithms for sorting the raw data into classifications: a plurality algorithm and majority algorithm. By the plurality algorithm, the classification with the most votes becomes the official classification.   By the majority algorithm, the official classification is one with $\ge 50$\% of the votes, if one exists.

To help test these algorithms, we prepared a ``gold standard'' data set of 500 subjects classified by the members of the science team.  All of the subjects in this ``gold standard'' data set were classified by two or more science team members, and all tie votes were discarded.  This process showed that science team members agree with one another at roughly the 82\% level as to whether an object is ``good'' or not.  A challenge in creating this "gold standard" set was that sometimes
science team members ruled out candidates based on their expertise/outside information rather than strictly
doing what the website asked for.  There was also a range in terms of how conservative the science team members were in terms of crowded fields and ISM background.  For example, the YSO experts (used to looking in the Galactic plane) on the science team tolerated background objects and "extended" objects more readily than the debris disk
experts on the team.

The majority algorithm performed well at selecting objects deemed to be classified as ``None of the Above/Good candidate''. On a randomly selected group of about 500 subjects, 94\% of those classified as ``None of the Above/Good candidate'' by the majority algorithm agreed with the gold standard set rankings. For the other five classification buttons, the agreement between classifications selected via the plurality algorithm and the gold standard set varied: 52\% for ``Extended'', 54\% for ``Not round'', 24\% for ``Object Moves Off the Crosshairs'', and 7\% for ``Empty circle''.  This lack of agreement suggests that users may be sometimes neglect to click {\it all} of the relevant buttons when a subject is ``bad'' for more than one reason. However, since our desire is simply to rule out false positives, this lack of agreement on the precise nature of the false positive does not hamper our study.

Deciding on classifications via the plurality algorithm generally leads to larger disagreements with the gold standard set and the classification data.  The exception to this rule seems to be the case of ``multiple objects in the red circle''; subjects classified as such by the plurality vote agree with the gold standard set 96\% of the time. 

Table~\ref{tab:buttons} shows the total selection distribution between the categories on the main site from the first $\sim 1$ million classifications. Users selected the ``good'' button nearly 20\% of the time.  Using the majority algorithm to determine a final classification, we get an 16\% yield of ``good'' objects based on those classifications.

Some of the brighter Disk Detective subjects show diffraction spikes and noise from detector saturation, especially in DSS, Sloan, and WISE 1 images.  When we launched Disk Detective we did not explicitly explain these phenomena anywhere on the site, though we readily answered questions about them in the chat forum TALK.  Nonetheless, many of our first users interpreted bright stars with spikes as ``oval'' or in some cases ``extended''.  On March 31, 2014, we edited the spotters guide and the tutorial at DiskDetective.org to include examples with diffraction spikes, which greatly reduced the problem, but some of these mistakes persist in our data. We expect this confusion over diffraction spikes to get better over time, as most of our classifications now come from participants who are well aware of the problem.

\subsection{Disk Detective Objects of Interest (DDOIs)}
\label{secondvetting}

With nearly 300,000 total objects from AllWISE to classify and 18-25\% yield described above, we estimate that the online classification scheme at DiskDetective.org will produce up to 75,000 good objects, still a daunting number to investigate.  However, the automated online classification stage is just the beginning of our vetting process.  The next stage of the process aims to harvest Disk Detective Objects of Interest (DDOIs), subjects that we consider deserving of additional follow-up observations. 

For this next stage of vetting, we created a collection of Google spreadsheets which both the science team and the superusers could edit.  We first populated the spreadsheets with subjects chosen as ``None of the Above/Good Candidate'' using the latest classification data. We chose subjects in the right portion of the sky for any upcoming follow-up observations and also populated the lists in order of agreement fractions and brightness (in J band), ensuring that the bright subjects, and those with high agreement percentages were looked at first.  Then we invited the superusers to add their own favorite objects, which they collect using tools on the TALK social network. We also invited the entire Disk Detective user community to submit subjects automatically via a Google form (but not directly edit the spreadsheet).

We coached the superusers on how to research each source in SIMBAD and VizieR (and sometimes NED) to fill in information about spectral type, proper motion, variability, parallax, prior observations, and make comments on the SED, etc. Then we checked the superusers's comments and selected the follow-up targets (DDOIs) from the list based on the following criteria.

\begin{enumerate}
\item 
SIMBAD object descriptions excluding post-AGB stars, carbon stars, novae, Cepheids, cataclysmic variables, high-mass x-ray binaries, eclipsing binaries, galaxies, Active Galactic Nuclei, planetary nebulae, reflection nebulae, rotational variables, symbiotic stars, or Wolf-Rayet stars. Note that we did keep sources with SIMBAD object descriptions Shell Star, Orion Variable, and White Dwarf.
\item
No Long Period Variables (LPVs), SR+L, Slow Irregular Variables, Miras, Semi-regular Variables, Semiregular Pulsating Variables, or Carbon stars based on literature searches.  
\item
Including only spectral types B through M according to SIMBAD, when a type is available.  
\end{enumerate}

Only about half of the subjects have entries in SIMBAD.  So we often relied on VizieR to help us search the relevant literature.  Many of the subjects had unknown spectral types, and many were severely reddened.  So at this stage of the vetting process, we often simply labeled sources as ``late-type based on color.'' 

\begin{enumerate}
\setcounter{enumi}{3}
\item
No sign that the WISE 1 photometry drops out due to saturation, based on visual inspection of the SED.
\item
No known companions within $16''$, except for spectroscopic binaries. 
\item
For M stars and other subjects with $V-J > 1$,  we require $[W1] - [W4] > 0.9$.  The peak of the thermal emission from cool star photospheres may lie at long enough wavelengths that the Rayleigh-Jeans limit no longer accurately describes the photosphere's $[W1] - [W4] $ color even in the absence of circumstellar dust.  Hence, we impose this more stringent requirement of 0.9 magnitudes of 22 $\mu$m excess for these cool subjects.  For some red subjects, however, we find that the SED clearly curves upwards at W4; we do not exclude these subjects.
\end{enumerate}
In the process, we naturally rediscovered many known disks. So we added the following additional criterion:
\begin{enumerate}
\setcounter{enumi}{6}
\item 
No sources that have already been imaged by a pointed space mission (i.e. {\it Spitzer}, Herschel, HST) or 8-m class telescope (i.e. Keck, Gemini, VLT, Magellan) (based on the published literature), except those without good quality spectral types (SIMBAD quality C or higher).
\end{enumerate}

Our search includes many sources in and near the Galactic plane and many sources with no parallax measurements from Hipparcos.  But since these sources require extra care, we have added two additional criteria for the purpose of this paper:
\begin{enumerate}
\setcounter{enumi}{7}
\item
No sources within $5^\circ$ of the Galactic plane.
\end{enumerate}

\begin{enumerate}
\setcounter{enumi}{8}
\item Only sources with parallax measurements from Hipparcos. 
\end{enumerate}

Most of the subjects on the vetting spreadsheets do not meet these additional eight criteria. But sources that do meet all these criteria (plus the criteria in Table~\ref{tab:cuts}, of course) we label as DDOIs and place in our our queue for follow-up observations.  So far we have collected 770 DDOIs in total that have survived the above vetting by the science team and/or by multiple superusers.  Of these, 517 have classification histories via the main Disk Detective online classification tool.  The remaining DDOIs have minimal classification data, since they were submitted directly by volunteers, and some volunteers choose to flip through images on TALK rather than the main site.  

The yield at this stage varies greatly depending on how we rank the spreadsheets.  Higher galactic latitudes provide higher yields, as do sources that are brighter in WISE 4.  Also, as our users have become more educated, they have become better at selecting subjects to place on the spreadsheets, which raises the yield.  But as of now, the typical yield at this stage of vetting (DDOIs per source on the vetting spreadsheet) is about 12\%. 

Multiplying the size of the input catalog by the automated vetting yield (24.4\%) and then by the DDOI vetting yield (about 12\%) gives a final yield of about 3\% and an estimate of the total number of DDOIs we ultimately expect to discover of $\sim 8000$. This estimate suggests that our search is presently about 10\% complete. However, it is worth noting that many of these criteria, while catching a lot of the aforementioned AGB stars that pollute our ``good'' candidate list, also eliminate a population of YSO disks and M stars (for example, YSOs can be irregular variables). 

\subsection{Using DDOIs As Quality Indicators}
\label{ddoisquality}

The DDOIs have all been carefully vetted by hand and researched in the literature by multiple scientists and/or well-trained superusers.  So this list of sources serves as a reference set of subjects that we can use to make decisions about how to run the vetting. 

\begin{figure*}
\begin{center}
\includegraphics[ totalheight=.6\textheight, angle=270]{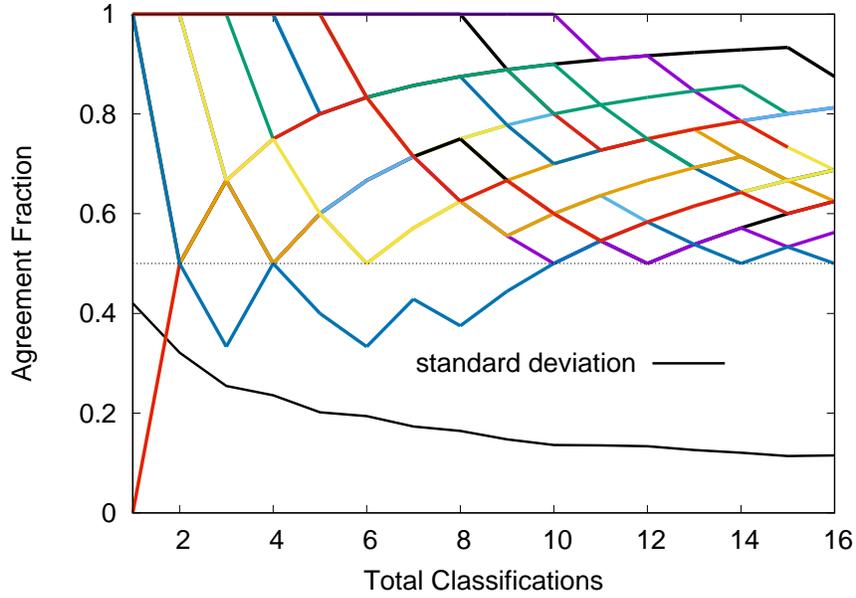}
\caption{User agreement and standard deviation of the agreement versus number of total classifications for DDOI candidates ultimately marked as ``good''. The faint lines show how the running agreement for individual subjects in the set vary with the number of classifications and the dark bold line shows shows the running standard deviation of the entire set. The standard deviation flattens out between 10-15 classifications, so we use 15 classifications as our cut-off for the retirement of subjects from the site.  \label{Fig_agreement}}
\end{center}
\end{figure*}

One major consideration of any citizen science project is when to remove, or ``retire,'' subjects from classifications on the site. To make that decision, we need to know roughly how long it takes the user population to converge on an answer.  Figure~\ref{Fig_agreement} shows how the standard deviation of the agreement for majority-algorithm ruled ``good'' candidates in our DDOI set varies with number of total classifications. For these subjects, ``agreement'' is defined simply as the number of good classifications divided by the total number of classifications the subject has received. Between 10-15 classifications, the standard deviation of the agreement for this subset of ``good'' DDOI candidates levels off, suggesting that there would be minimal marginal benefit from requiring additional classifications beyond this point. Thus the current iteration of the site uses a conservative 15 total classifications as our benchmark for when to retire subjects.

\section{Follow-Up Observations}
\label{sec:followup}

We obtained and analyzed spectra for our DDOIs using the FAST spectrograph \citep{1998PASP..110...79F} on the Tillinghast 1.5m telescope at Fred Lawrence Whipple Observatory during May--October 2014, employing the 300 mm${}^{-1}$ grating and the 3'' slit.  These spectra cover 3800--7500 \AA~at
a resolution of $\sim$ 6 \AA.
We flux- and wavelength-calibrated the spectra using the Image Reduction and Analysis Facility (IRAF) software system.  After trimming the CCD frames at each end of the slit,
we corrected for the bias level, flat-fielded each frame,
applied an illumination correction, and derived a full wavelength
solution from calibration lamps acquired immediately after each
exposure.  The wavelength solution for each frame has a probable
error of $\pm$0.5--1.0 \AA.  We extracted sources and sky spectra
using the optimal extraction algorithm within APEXTRACT.  The
absolute flux-calibration for each night relies on observations
of 2-5 standard stars \citep{1975ApJ...197..593H, bar82, 1988ApJ...328..315M} and has 
an uncertainty of $\pm$5\%--10\%.

This follow-up spectroscopy has proven vital to our vetting of DDOIs.  Roughly half of all DDOIs initially have no reliable spectral type. Additionally, for these objects the luminosity class is generally completely unconstrained, save for clues from parallax and proper motion measurements.  This situation is more dire for red (or reddened) sources and late-type stars, which will be the subject of future work. M giants often produce their own dust, so disks around these stars are not of particular interest from the perspective of planet formation.  FAST spectroscopy also allows us to screen for certain false positives, such as blended AGN.  

Though our initial candidate list in this paper is small, we plan to follow-up on the entire list of DDOIs and thus will have a large number of observed spectra. Manual spectral classification of every object is highly inefficient. To speed up the process, we used the semiautomatic quantitative spectral-typing code SPTCLASS\footnote{ http://dept.astro.lsa.umich.edu/~hernandj/SPTCLASS/sptclass.html}, an IRAF/IDL code based on the methodology outlined in \citet{2004AJ....127.1682H}. We decided that in this first paper we would only publish DDOIs that are in Hipparcos catalogs, and consequently do
have spectral types inferred from color and parallax, as a test of our pipeline. 

SPTCLASS calculates spectral types of stars using spectral indices, comparing line fluxes of spectral features that are sensitive to effective temperature \citep[e.g.,][and others]{1943assw.book.....M, 1999RMxAA..35..143S, 1999yCat.6071....0C, 2001AJ....121.2148G, 1977ApJS...34..101P, 1995AJ....110.1838R}. SPTCLASS uses three independent spectral typing modules: indices characterizing early (OBA, 44 indices), intermediate (FG, 11 indices), and late (KM, 16 indices) spectral types. Each index is based on the equivalent width for the spectral feature, which is calculated by measuring the decrease in flux from the expected continuum due to line absorption. Indices measured by this procedure are generally insensitive to reddening, so long as each band's wavelength coverage is relatively small. The indices have been calibrated as a function of spectral type using O8-M6 main sequence standards; this large extent for all indices ensures that degeneracy of determination is eliminated. 

SPTCLASS first calculates three spectral types for each star, one for each module, by taking a weighted average of the indices used in that module, using weights estimated from the computed error for each index.  It then calculates a weighted average of spectral types for all indices, using weights estimated by the computed error for each index from the measured index value and fit of the index to spectral type. The code then discards spectral indices if the spectral type they indicate deviates from the mean by $>3 \sigma$, or if their computed error is more than 6 subtypes. This minimizes possible contamination by artifacts and emission lines.  We report here the final SPTCLASS spectral type, which is the new average after discarding these deviant indices.  The typical 1-$\sigma$ formal uncertainty for these classifications is 2 subtypes.

\subsection{First Batch of Disk Candidates}
\label{sec:candidates}

Table~\ref{tab:ddois} lists our first batch of 50 DDOIs.  Although some of the sources in Table~\ref{tab:ddois} are Be stars or disk candidates that have been previously reported, 37 of these sources are new disk candidates from the Disk Detective project.  

We have decided to present only stars in with Hipparcos parallaxes as a test of our spectral classification pipeline.  Though \citet{2013ApJS..208...29W} performed a large search for stars in the Hipparcos catalog that have excess emission in W4, they used a $[K_s] - [W4]$ cutoff, and the AllSky catalog.  Our selection criteria $[W1] - [W4]$ and input catalogs (AllWISE) are different, so we find objects they missed, even in the Hipparcos catalog.

Table~\ref{tab:ddois} lists the photometry of these disk candidates from 2MASS and SIMBAD (V band).  It also lists the spectral types we derived from our FAST spectra, informed by photometry and Hipparcos distances, and WISE photometry.
Eleven of the sources in Table~\ref{tab:ddois} have previously been reported as disk candidates by 
\citet{2012ApJ...752...58Z}, \citet{2012MNRAS.426...91K}, \citet{2013MNRAS.433.2334K}, \citet{2013ApJS..208...29W}, \citet{2013ApJ...773..179W} and \citet{2014ApJS..211...25C}.  We include these objects here as a check on our consistency with other searches. However, 37 of these sources have not been previously reported as disk candidates (and are not Be stars).

The WISE photometry in Table~\ref{tab:ddois} is taken from the All-Sky data release and corrected for saturation according to the formulae in \citet{2014ApJS..212...10P}.   This corrected All-Sky WISE photometry is more accurate for sources brighter than about 8th magnitude in W1.   No color correction is applied to the WISE photometry, since the sources have nearly Rayleigh-Jeans spectra.  If we assumed instead that the emission were entirely from a 200 K blackbody disk, the correction to WISE 4 would be  -0.016 magnitudes \citep{2010AJ....140.1868W}, negligible compared to our 0.25 magnitude WISE 1-WISE 4 excess criterion for inclusion in Disk Detective.

\begin{figure*}
\begin{center}
\includegraphics[ totalheight=.6\textheight, angle=270]{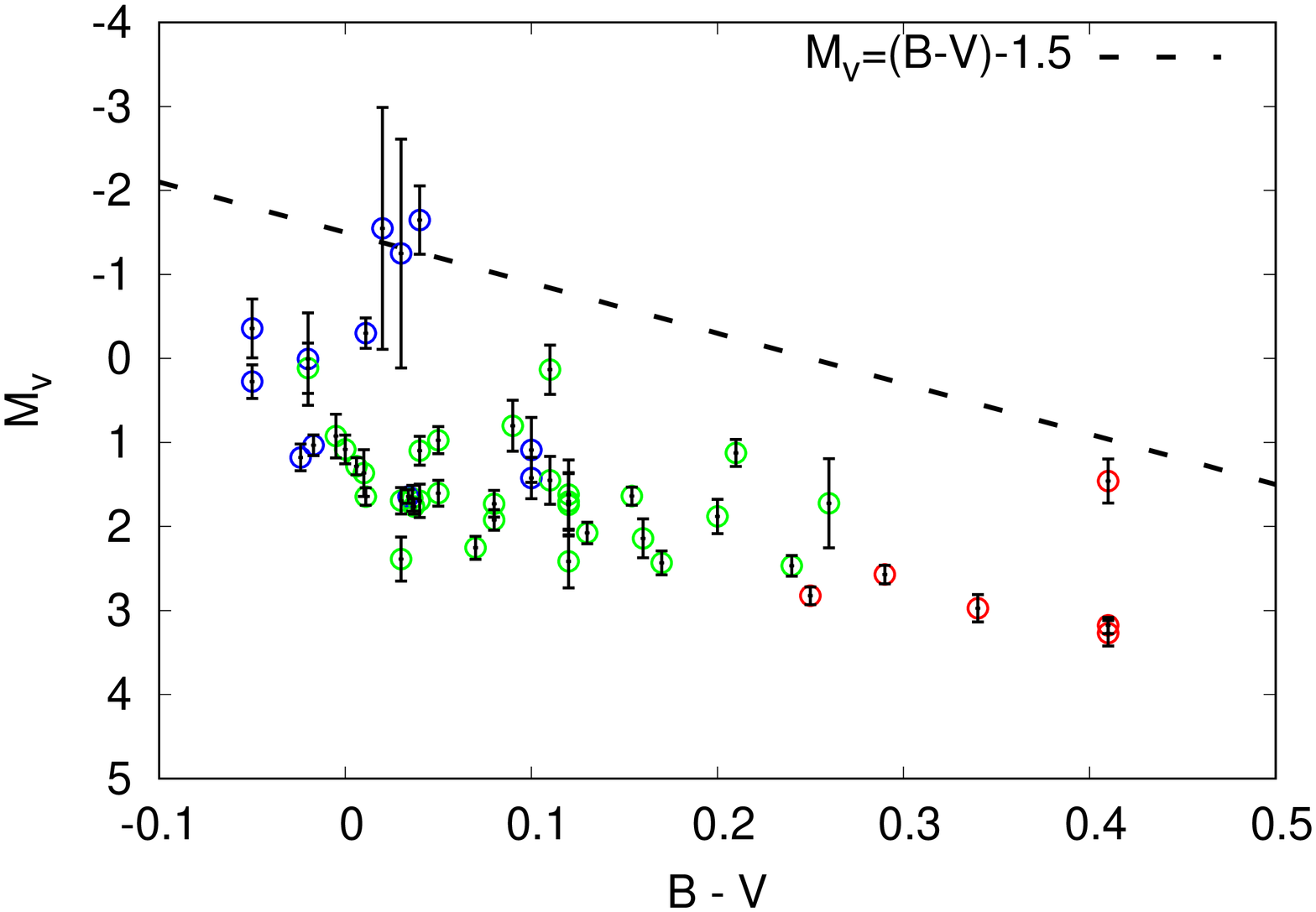}
\caption{Most disk candidates in this paper easily fall below the limit on the HR diagram used by Patel et al (2014) to separate giants from dwarfs. The three stars that lie above this limit (HD 7406, HD 164137, and HD 2185461) are Be stars. The points are color coded according to spectral type: blue=B, green=A, red=F.  \label{Fig_HR}}
\end{center}
\end{figure*}

The vast majority of our DDOIs did not have previously-identified luminosity classes.  We assigned them using the approach of
 \citet{2014ApJS..212...10P}, who separated dwarfs from giants in their catalog via a simple cut on an HR diagram, retaining only stars with $M_V > 6.0(B-V)-1.5$.  Fig.~\ref{Fig_HR} shows an HR diagram for our candidates.  The stars on Fig.~\ref{Fig_HR} are color coded according to spectral type: blue=B, green=A, red=F.  Most of our disk candidates are probably dwarfs since they fall below the  \citet{2014ApJS..212...10P} cut, which is shown by the dashed line.

\begin{deluxetable}{lllcccccl}
\rotate
\tabletypesize{\footnotesize}
\tablecolumns{9}
\tablewidth{0pt}
\tablecaption{Observational Parameters for Disk Detective Objects of Interest}
\tablehead{\multicolumn{3}{c}{Identifiers} &
	\colhead{Spectral} &
	\colhead{Distance} &
	\multicolumn{2}{c}{Photometry} &
	\colhead{Excess} &
	\colhead{} \\
	\colhead{Zooniverse} &
	\colhead{HD Number} &
	\colhead{WISE} &
	\colhead{Type} &
	\colhead{(pc)} &
	\colhead{V\tablenotemark{1}} &
	\colhead{J} &
	\colhead{(W1-W4)} &
	\colhead{Notes} }

\startdata
	AWI00055sz  & 7406   &  J011636.23+740136.6   &  B1III    & $549   \pm 103$   & 7.05  & $6.764 \pm 0.056$ & $0.619 \pm 0.052$ &  a, EW$= -17.7064$ \\
	AWI0005w41  & 218546 &  J230817.21+511146.3   &  B8III    & $794   \pm 498$   & 8.25  & $7.918 \pm 0.020$ & $0.784 \pm 0.072$ &  \\
	AWI0005bow  & 164137 &  J175911.27+135417.8   &  B8III    & $820   \pm 544$   & 8.02  & $7.768 \pm 0.020$ & $1.571 \pm 0.050$ &  \\
	AWI0005ae4  & 224098 &  J235449.26+742436.2   &  B8V      & $240   \pm 20 $   & 6.6   & $6.388 \pm 0.019$ & $0.433 \pm 0.050$ &  \\
	AWI0000m2p  & 4670   &  J004848.00+181850.6   &  B9IV     & $386   \pm 98 $   & 7.94  & $7.841 \pm 0.020$ & $0.898 \pm 0.080$ &  \\
	AWI00055sx  & 6370   &  J010652.55+743754.5   &  B9IV     & $286   \pm 51 $   & 8.37  & $8.168 \pm 0.029$ & $0.919 \pm 0.084$ &  \\
	AWI0005yiz  & 6612   &  J010722.60+380143.9   &  B9V      & $316   \pm 51 $   & 7.14  & $7.156 \pm 0.020$ & $0.453 \pm 0.067$ &  b, EW$= 4.09265$ \\
	AWI00062a8  & 9985   &  J013756.15+211539.9   &  B9V      & $202   \pm 23 $   & 7.95  & $7.824 \pm 0.018$ & $1.993 \pm 0.050$ &  \\
	AWI0005ym7  & 23873  &  J034921.76+242251.0   &  B9V      & $122   \pm 9  $   & 6.61  & $6.602 \pm 0.017$ & $1.138 \pm 0.057$ &  \\
	AWI0000uj2  & 138422 &  J153046.05+342756.4   &  B9V      & $108   \pm 4  $   & 6.81  & $6.719 \pm 0.021$ & $0.542 \pm 0.044$ & c \\
	AWI00004o8  & 152308 &  J165204.85+145827.2   &  B9V      & $124   \pm 7  $   & 6.5   & $6.504 \pm 0.034$ & $0.494 \pm 0.048$ &  \\
	AWI00062fh  & 207888 &  J215221.25-031028.9   &  B9V      & $184   \pm 17 $   & 6.6   & $6.688 \pm 0.026$ & $1.235 \pm 0.045$ &  \\
	AWI0000bs0  & 2830   &  J003140.76-014737.3   &  A0V      & $107   \pm 6  $   & 7.07  & $6.917 \pm 0.019$ & $1.015 \pm 0.054$ & c \\
	AWI0004ne5  & 9590   &  J013525.89+560237.3   &  A0V      & $162   \pm 12 $   & 7.02  & $6.791 \pm 0.018$ & $0.655 \pm 0.059$ &  \\
	AWI0005yjp  & 14893  &  J022515.75+370707.9   &  A0V      & $192   \pm 23 $   & 7.34  & $7.341 \pm 0.018$ & $1.037 \pm 0.067$ &  \\
	AWI0005ylw  & 22614  &  J033906.73+244209.8   &  A0V      & $120   \pm 11 $   & 7.09  & $6.979 \pm 0.024$ & $0.614 \pm 0.079$ &  \\
	AWI00000wz & 74389  &  J084546.93+485243.4   &  A0V      & $111   \pm 7   $   & 7.48  & $7.296 \pm 0.023$ & $0.678 \pm 0.067$ & \\
	AWI0000u8s  & 85672  &  J095359.12+274143.5   &  A0V      & $107   \pm 7  $   & 7.58  & $7.215 \pm 0.024$ & $1.011 \pm 0.056$ &  c, d \\
	AWI0000w9x  & 129584 &  J144313.04+014928.7   &  A0V      & $156   \pm 20 $   & 7.33  & $7.260 \pm 0.024$ & $0.785 \pm 0.052$ &  \\
	AWI0000tz1  & 140101 &  J154030.20+370101.1   &  A0V      & $165   \pm 13 $   & 7.17  & $7.109 \pm 0.023$ & $0.688 \pm 0.051$ &  c \\
	AWI0000gjb  & 214982 &  J224206.62-032824.4   &  A0V      & $124   \pm 9  $   & 7.16  & $7.101 \pm 0.021$ & $0.655 \pm 0.065$ &  \\
	AWI0000kg4  & 218155 &  J230533.05+145732.5   &  A0V      & $106   \pm 5  $   & 6.77  & $6.712 \pm 0.020$ & $0.579 \pm 0.059$ &  \\
	AWI0000fye  & 224155 &  J235537.71+081323.7   &  A0V      & $128   \pm 6  $   & 6.82  & $6.770 \pm 0.026$ & $0.462 \pm 0.066$ &  \\
	AWI00062h1  & 224429 &  J235746.21+112827.6   &  A0V      & $95    \pm 4  $   & 6.65  & $6.581 \pm 0.023$ & $0.426 \pm 0.058$ &  c \\
	AWI0002mhd  & 173056 &  J184305.97+071626.4   &  A0.5V    & $203   \pm 32 $   & 8.24  & $7.913 \pm 0.027$ & $1.504 \pm 0.061$ &  \\
	AWI0005yjn  & 14685  &  J022317.32+381509.7   &  A1IV     & $254   \pm 35 $   & 7.14  & $7.037 \pm 0.018$ & $0.706 \pm 0.062$ &  e, EW$= 4.01698$ \\
	AWI0005abf  & 213290 &  J222753.27+704806.0   &  A1IV     & $326   \pm 44 $   & 7.7   & $7.318 \pm 0.023$ & $0.627 \pm 0.050$ &  \\
	AWI0005mry  & 3051   &  J003412.66+540359.0   &  A1V      & $229   \pm 32 $   & 7.6   & $7.350 \pm 0.019$ & $0.354 \pm 0.073$ &  \\
	AWI0000phh  & 18271  &  J025614.05+040254.2   &  A1V      & $116   \pm 14 $   & 7.71  & $7.664 \pm 0.032$ & $0.974 \pm 0.073$ &  \\
	AWI0005zz5  & 25466  &  J040238.47-004803.7   &  A1V      & $113   \pm 8  $   & 6.93  & $6.835 \pm 0.020$ & $0.606 \pm 0.053$ &  \\
	AWI0000uji  & 134854 &  J151147.67+101259.8   &  A1V      & $113   \pm 8  $   & 6.87  & $6.799 \pm 0.018$ & $0.405 \pm 0.065$ &  \\
	AWI0005zx4  & 11085  &  J014928.21+244048.7   &  A2V      & $151   \pm 22 $   & 8.31  & $8.013 \pm 0.019$ & $0.641 \pm 0.088$ &  \\
	AWI0004nfu  & 21375  &  J032853.67+490412.8   &  A2V      & $160   \pm 21 $   & 7.47  & $7.186 \pm 0.026$ & $0.468 \pm 0.068$ & e \\
	AWI00062aj  & 12445  &  J020221.16+192323.6   &  A3V      & $227   \pm 43 $   & 8.4   & $8.090 \pm 0.024$ & $0.944 \pm 0.084$ &  \\
	AWI0000tgc  & 84870  &  J094902.82+340506.9   &  A3V      & $88    \pm 5  $   & 7.19  & $6.752 \pm 0.032$ & $0.398 \pm 0.063$ & c, f \\
	AWI0005bps  & 165507 &  J180533.55+182643.9   &  A3V      & $193   \pm 33 $   & 8.17  & $7.952 \pm 0.019$ & $0.672 \pm 0.099$ &  \\
	AWI0005vyx  & 204829 &  J212959.78+413037.3   &  A3V      & $175   \pm 13 $   & 7.34  & $6.744 \pm 0.021$ & $0.338 \pm 0.067$ &  \\
	AWI0005w29  & 212556 &  J222412.79+484918.7   &  A3V      & $150   \pm 11 $   & 7.61  & $7.415 \pm 0.029$ & $0.469 \pm 0.060$ &  \\
	AWI0005a9r  & 208410 &  J215305.45+682955.0   &  A3V      & $189   \pm 15 $   & 7.48  & $7.297 \pm 0.027$ & $0.539 \pm 0.053$ &  \\
	AWI0005ykd  & 21062  &  J032448.99+283908.6   &  A4V      & $102   \pm 6  $   & 7.12  & $6.838 \pm 0.018$ & $0.612 \pm 0.058$ &  \\
	AWI0000v1z  & 138214 &  J152954.11+234901.6   &  A5V      & $138   \pm 13 $   & 7.58  & $7.124 \pm 0.024$ & $0.322 \pm 0.061$ &  \\
	AWI0006222  & 201377 &  J210916.04-001405.6   &  A7V      & $101   \pm 5  $   & 6.66  & $6.355 \pm 0.032$ & $0.574 \pm 0.064$ &  \\
	AWI0005w7o  & 20994  &  J032429.84+341709.9   &  A8V      & $245   \pm 60 $   & 8.67  & $7.931 \pm 0.027$ & $1.191 \pm 0.074$ &  \\
	AWI0005c01  & 199392 &  J205143.50+730449.3   &  A8V      & $169   \pm 18 $   & 8.28  & $8.004 \pm 0.023$ & $0.402 \pm 0.071$ &  \\
	AWI00002ms  & 71988  &  J083100.44+185806.0   &  F0V      & $83    \pm 4  $   & 7.42  & $6.977 \pm 0.032$ & $0.358 \pm 0.080$ &  \\
	AWI00062bl  & 22128  &  J033337.91-072453.8   &  F0IV     & $166   \pm 20 $   & 7.56  & $6.919 \pm 0.024$ & $0.853 \pm 0.053$ &  \\
	AWI0005yk3  & 19257  &  J030651.95+303136.8   &  F0V      & $79    \pm 4  $   & 7.06  & $6.470 \pm 0.020$ & $2.093 \pm 0.037$ & g \\
	AWI000048c  & 87827  &  J100719.80-152718.9   &  F2V      & $107   \pm 8  $   & 8.12  & $7.425 \pm 0.020$ & $0.968 \pm 0.060$ & c \\
	AWI0000hjr  & 221853 &  J233536.20+082256.9   &  F2V      & $68    \pm 3  $   & 7.34  & $6.559 \pm 0.019$ & $1.335 \pm 0.042$ & c,f,h \\
	AWI00002yt  & 157165 &  J172007.53+354103.6   &  F6V      & $100   \pm 7  $   & 8.27  & $7.496 \pm 0.020$ & $0.427 \pm 0.064$ &  \\
\enddata

\tablenotetext{1}{From SIMBAD. Typical uncertainty 0.01 mag.}
\tablenotetext{a}{H$\alpha$ partially in emission.}
\tablenotetext{b}{H$\alpha$ in emission.}
\tablenotetext{c}{Appears in Wu et al. (2013).}
\tablenotetext{d}{Appears in Wahhaj et al. (2013).}
\tablenotetext{e}{Appears in Zuckerman et al. (2012).}
\tablenotetext{f}{Appears in Chen et al. (2014).}
\tablenotetext{g}{Appears in Kennedy \& Wyatt (2013).}
\tablenotetext{h}{Appears in Kennedy \& Wyatt (2014).}

  \label{tab:ddois}

\end{deluxetable}

If we conservatively assume that the quality of our spectra corresponds to the ``C'' quality flag in SIMBAD (``A'' is the highest quality), then of the 50 targets with new spectral types that we present here, 48 show clear improvements in quality over the published literature, while the remaining nine targets have spectral types of the same quality as the literature. The root mean square change in spectral types versus SIMBAD for all 50 targets was 2.64, with 27 objects shifted at least one subtype toward lower temperature and 12 objects shifted at least one subtype toward higher temperature.

\begin{figure*}
\begin{center}
\includegraphics[ totalheight=.6\textheight, angle=0]{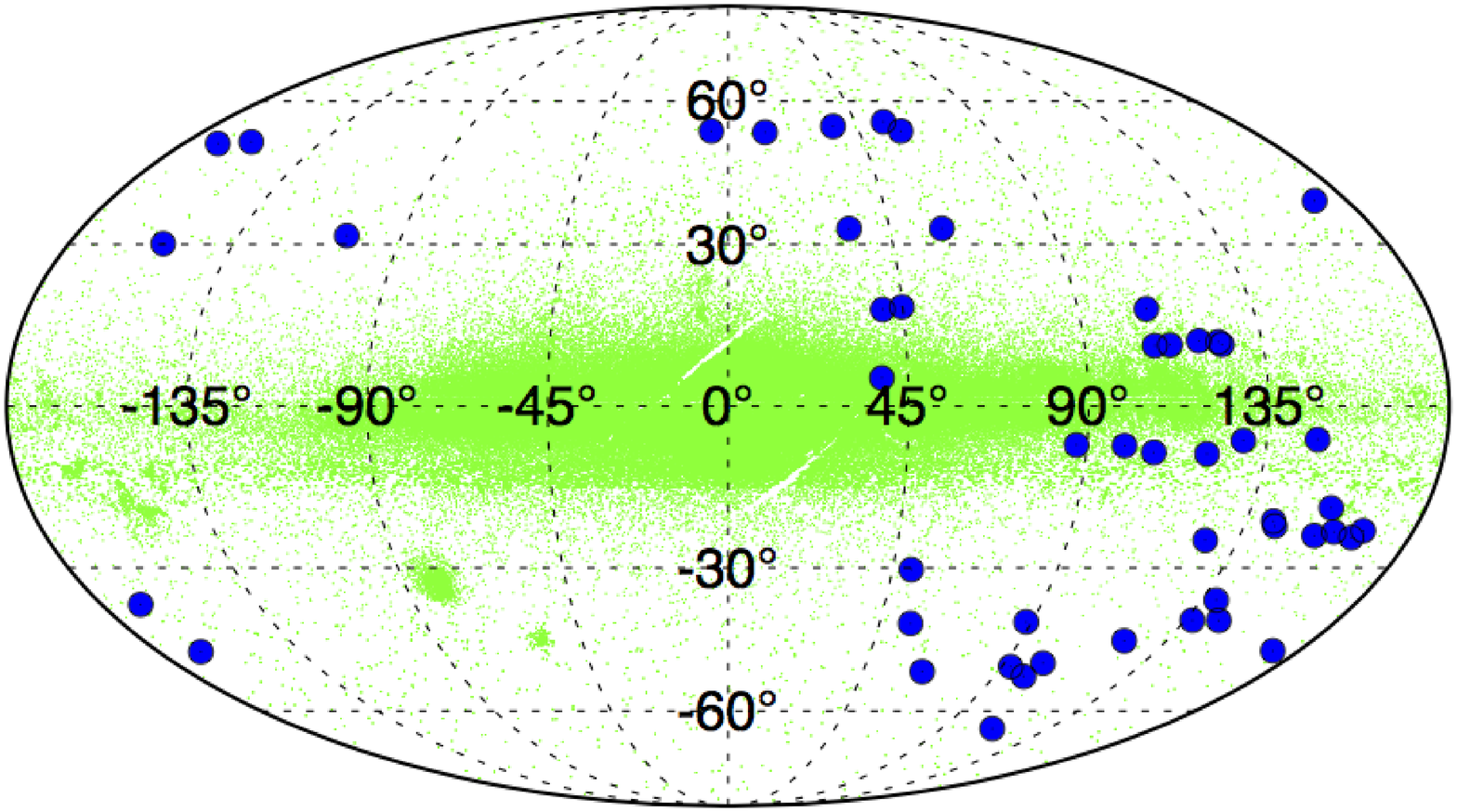}
\caption{Galactic latitude and longitude distribution of the disk candidates in Table 1 (blue points) and all of the subjects pre-selected based on the criteria in Table 1 plus the J$<$14.5 criterion (green points).  The distribution of the DDOIs on the sky is shaped mainly by the range of declinations accessible to FAST and our decision to add sources to the website grouped by Galactic latitude.  \label{Fig_radec}}
\end{center}
\end{figure*}

\begin{figure*}
\begin{center}
\includegraphics[ totalheight=.6\textheight, angle=270]{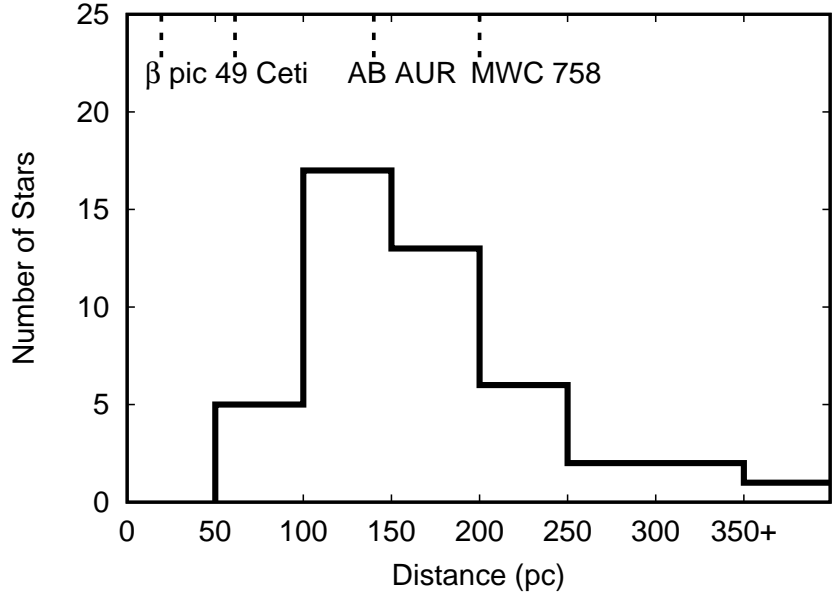}
\caption{Distribution of distances to the DDOIs listed in Table~\ref{tab:ddois} from Hipparcos parallax measurements.  
The distance distribution of the DDOIs is grossly similar to that of the disk candidates in \citet{2013ApJS..208...29W}, but also includes a long tail of objects beyond 200 pc, which consists mainly of B and early A dwarfs.
\label{Fig_dist}}
\end{center}
\end{figure*}

Figures~\ref{Fig_radec} and \ref{Fig_dist} summarize some more properties of the stars in Table~\ref{tab:ddois}: their distribution on the sky and their distance distribution.  Their distance distribution peaks at roughly 110 pc; at this distance, a large telescope with adaptive optics like the Gemini Planet Imager could image an analog to the HR 8799 planetary system and comfortably detect at least two of the planets.  For comparison, \citet{2014ApJS..212...10P} limited their study to sources with distances < 120 pc.  But the distance distribution of the DDOIs is grossly similar to that of the disk candidates in \citet{2013ApJS..208...29W}, which peaks at about 90 pc.  The distribution of our DDOIs also includes a long tail of objects beyond 200 pc; roughly 1/4 of our DDOIs are in this long tail, which consists mainly of B and early A dwarfs.  Fig.~\ref{Fig_dist} also shows the distances to some well-known disks for references.  The distribution of the DDOIs on the sky is shaped mainly by the range of declinations accessible to FAST and our decision to add sources to the website grouped by Galactic latitude. 

\begin{figure*}
\begin{center}
\includegraphics[ totalheight=.6\textheight, angle=270]{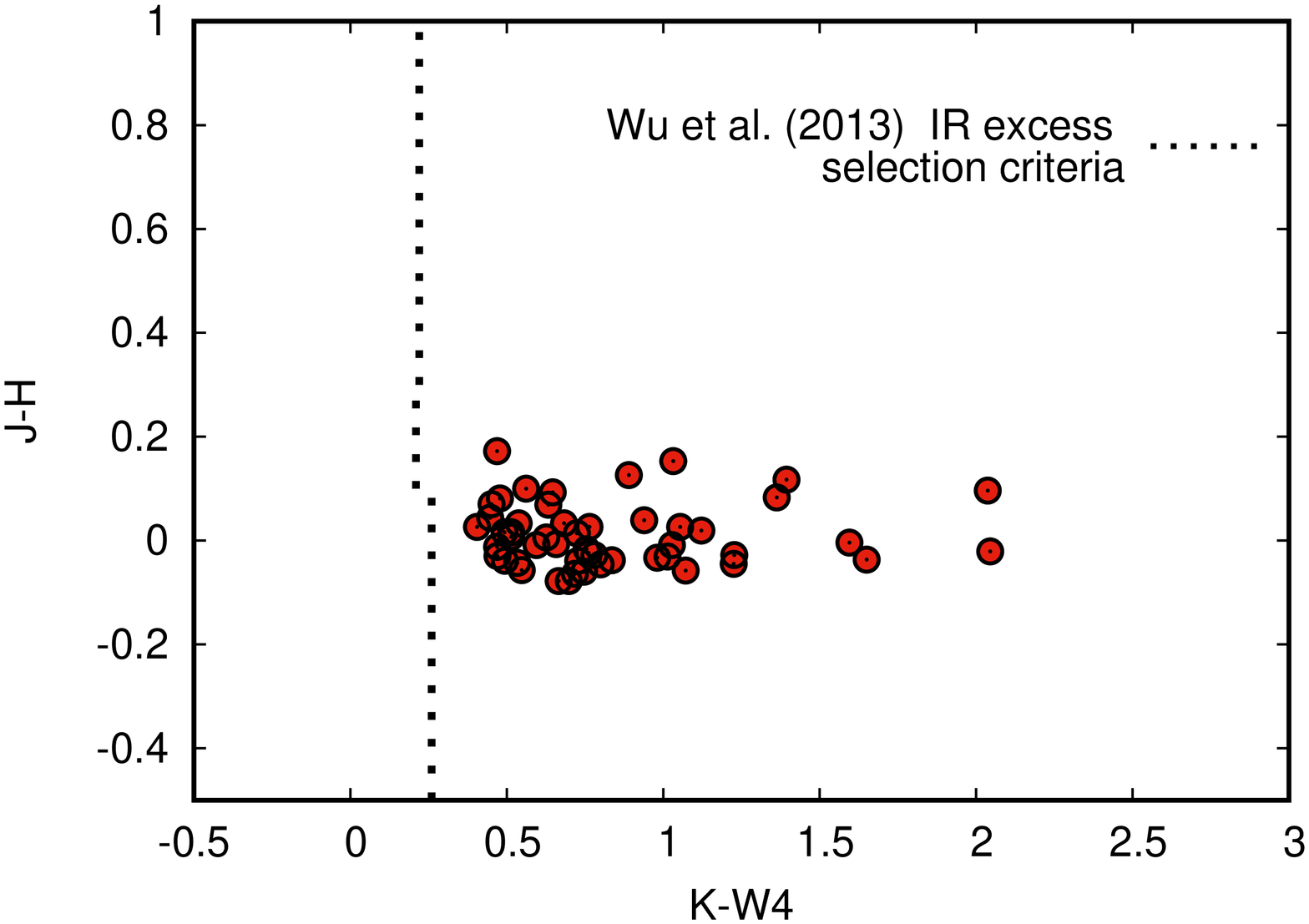}
\caption{J-H versus K-W4 for the sources presented in Table~\ref{tab:ddois}. J,H,K magnitudes are from 2MASS, and W4 is the 22 $\mu m$ magnitude from AllWISE. The faint dashed line shows the criterion used by \citet{2013ApJS..208...29W} for selecting disk candidates with 22  $\mu m$ excess (see their Figure 2). Even though we selected our subjects based on W1-W4, our disk candidates all satisfy their criterion as well.  \label{Fig_Wu}}
\end{center}
\end{figure*}

Figure~\ref{Fig_Wu} compares our disk candidates with the \citet{2013ApJS..208...29W} color selection criteria.  We did not impose such a criterion, but all of our disk candidates fall to the right of the dotted line in this figure, showing that they meet the \citet{2013ApJS..208...29W} criterion anyway.  On average, the disk candidates in this paper are somewhat redder than those in Wu et al. (2013). 

We fit simple models to the SEDs consisting of a stellar photosphere plus a blackbody dust component using a Levenberg-Marquardt algorithm.  These fits yield constraints on the dust temperature and the fractional infrared luminosity, $f$, the total bolometric power emitted by our single-temperature blackbody disk model divided by the total bolometric power emitted by our blackbody stellar model for each source.  For stars with excess at only WISE 4, these fits yield only an upper limit on the dust temperature and a lower limit on $f$.

\begin{figure*} 
\begin{center}
\includegraphics[ totalheight=.6\textheight, angle=0]{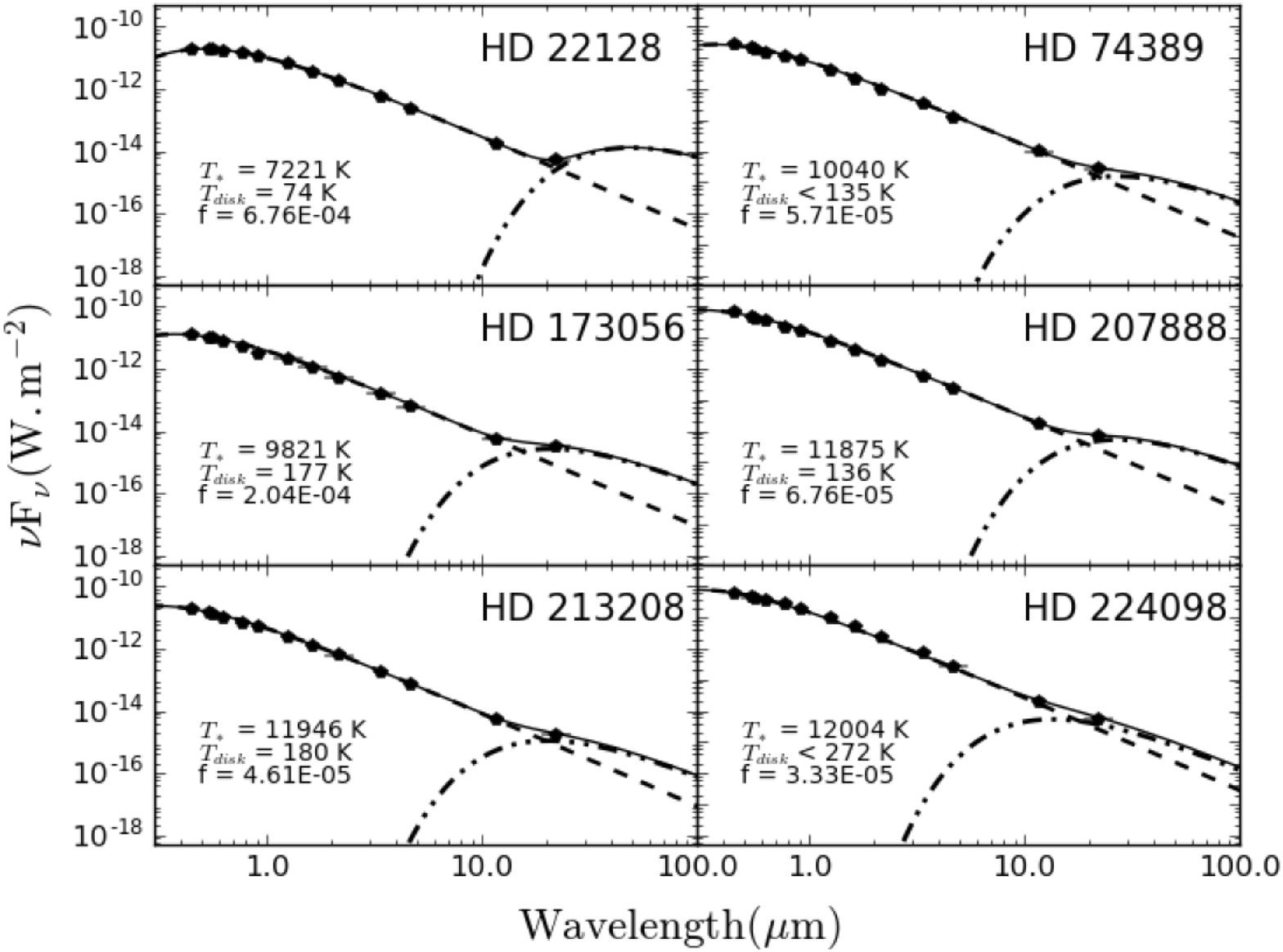}
\caption{Sample spectral energy distributions (SEDs) for six new disk candidates. Dashed lines show blackbody models fit to the star and to the disk component; the solid line shows the total model flux. \label{Fig_SEDs}}
\end{center}
\end{figure*}

Figure~\ref{Fig_SEDs} shows spectral energy distributions (SEDs) for six of the new disk candidates, together with a simple model for the flux.  Our SEDs employ WISE 1 and WISE 2 fluxes from the All-Sky survey catalog corrected for saturation using the formulae in \citet{2014ApJS..212...10P}.  For these models, we supplemented the WISE and 2MASS photometry with UBVRI photometry from SIMBAD when it was available. Dashed lines show blackbody models fit to the star and to the disk component; the solid line shows the total model flux. 

Table~\ref{tab:diskcandidates} summarizes our SED modeling of all the 39 new disk candidates reported in Table~\ref{tab:ddois}.  The temperature uncertainties are $1-\sigma$ uncertainties from the shape of the two-parameter $\chi^2$ surface near the minimum.  The fractional infrared excesses, $f$, listed in this table should all be considered lower limits.  When the temperature listed is an upper limit, the listed fractional infrared excess corresponds to a blackbody disk with the listed upper limit temperature.

\newcommand\sk[2]{Sk\,{$-#1{^\circ}#2$}}
\newcommand\tnc{\,\tablenotemark{c}}
\newcommand\tnd{\,\tablenotemark{d}}

\begin{centering}
\begin{deluxetable}{lccc}
\tabletypesize{\footnotesize}
\tablecolumns{4}
\tablewidth{0pt}
\tablecaption{Disk Parameters for New Disk Candidates}
\tablehead{\colhead{HD Number}	&
	\colhead{Disk Temperature (K)}	&
	\colhead{$f$}			}
\startdata
               3051              &          $ < 248         $                      &   <$4.7\times 10^{-5}$   \\
               4670              &           $ 156$        +21/-19             &     $ 5.0\times 10^{-5}$   \\   
               6370               &        $< 168    $                            &      <$6.9\times 10^{-5}$   \\
               6612              &           $< 189 $                              &     <$ 2.1\times 10^{-5}$   \\
               9590              &          $ < 200        $                      &      <$5.8\times 10^{-5}$   \\
               9985              &           $< 136  $                            &      <$2.8\times 10^{-4}$   \\
              11085              &          $<  200  $                            &     <$7.2\times 10^{-5}$   \\
              12445              &           $< 157  $                            &     <$8.9\times 10^{-5}$  \\
              14893             &           $ < 117  $                              &      <$3.4\times 10^{-5}$   \\
              18271             &           $ < 104  $                             &     <$8.7\times 10^{-5}$   \\
              20994              &          $  187  $    +25/-19                &     $2.5\times 10^{-4}$   \\   
              21062              &          $ < 207   $                             &    <$8.0\times 10^{-4}$   \\
              22128              &           $<  74     $                           &    <$6.8\times 10^{-4}$   \\
              22614             &          $ < 187         $                     &       <$6.8\times 10^{-5}$   \\
              23873             &            $< 131 $                             &       <$7.3\times 10^{-5}$   \\
              25466             &          $ < 107        $                       &      <$3.3\times 10^{-5}$   \\           
              71988              &         $  < 221          $                        &   <$1.1\times 10^{-4}$   \\
              74389             &          $ < 136      $                        &       <$5.7\times 10^{-5}$   \\
             129584            &           $ <  170     $                       &       <$4.2\times 10^{-5}$   \\
             134854            &           $ < 251       $                       &       <$4.6\times 10^{-5}$   \\
             138214             &          $ < 216        $                        &     <$9.6\times 10^{-5}$   \\
             152308            &             $< 228     $                         &        <$2.6\times 10^{-5}$   \\
             157165              &         $ < 245           $                     &     <$1.4\times 10^{-4}$    \\
             158419            &            $<  244 $                              &        <$4.1\times 10^{-5}$   \\
             165507             &          $  < 205         $                       &     <$7.2\times 10^{-5}$   \\
             173056            &           $   177        $   +20/-18     &       $2.0\times 10^{-4}$   \\    
             199392             &          $ < 252        $                       &      <$8.4\times 10^{-5}$   \\
             201377             &          $ < 131        $                       &      <$1.7\times 10^{-5}$   \\
             204829             &          $ < 263        $                       &      <$5.3\times 10^{-5}$   \\
             207888            &             $ <136        $                     &        <$6.8\times 10^{-5}$   \\
             208410             &          $  < 212         $                       &     <$7.4\times 10^{-5}$   \\
             212556             &          $ < 225        $                       &      <$6.2\times 10^{-5}$   \\
             213290            &          $  < 198      $                        &       <$7.3\times 10^{-5}$   \\
             214982            &          $  < 193      $                       &        <$6.3\times 10^{-5}$   \\
             218155            &          $ < 193      $                       &        <$6.3\times 10^{-5}$   \\
             224098            &         $  < 272    $                         &        <$3.3\times 10^{-5}$  \\
             224155            &           $ < 242      $                        &       <$4.4\times 10^{-5}$   \\
\enddata

\label{tab:diskcandidates}
\end{deluxetable}
\end{centering}

\subsection{W3 Excesses}

All of the diskdetective.org subjects were pre-selected to have [W1]-[W4] excesses significant at the 5-sigma level or better, based on the AllWISE catalog.  We also checked the All-Sky catalog of WISE photometry for these sources, correcting this photometry via the correction factors in \citet{2014ApJS..212...10P}.  All of the sources in Table~\ref{tab:diskcandidates} still have [W1]-[W4] excesses significant at the 5-sigma level or better using the corrected All-Sky photometry.

Additionally, eight of our DDOIs also turned out to have significant ($>3\sigma$) W3 excess based on corrected All-Sky WISE photometry, as you can read in Table~\ref{tab:w3}: HD 4670, HD 7406, HD 14685, HD 19257, HD 20994, HD 164137, HD 173056, and HD 218546.

\begin{deluxetable}{lllccc}
\tabletypesize{\footnotesize}
\tablecolumns{6}
\tablewidth{0pt}
\tablecaption{Disk Detective Objects of Interest with W3 Excess}
\tablehead{\colhead{Zooniverse ID} &
	\colhead{HD Number} &
	\colhead{WISE ID} &
	\colhead{Spectral Type} &
	\colhead{Distance} &
	\colhead{W1-W3}}

\startdata
	AWI00055sz  & 7406  &  J011636.23+740136.6   &  B1III    & $549   \pm 103$   & $0.292 \pm 0.042$ \\
	AWI0005w41  & 218546 &  J230817.21+511146.3   &  B8III    & $794   \pm 498$   & $0.102 \pm 0.029$ \\
	AWI0005bow  & 164137 &  J175911.27+135417.8   &  B8III    & $820   \pm 544$   & $0.646 \pm 0.029$ \\
	AWI0000m2p  & 4670  &  J004848.00+181850.6   &  B9IV     & $386   \pm 98$    & $0.115 \pm 0.030$ \\
	AWI0002mhd  & 173056 &  J184305.97+071626.4   &  A0.5V    & $203   \pm 32$    & $0.114 \pm 0.030$ \\
	AWI0005yjn  & 14685 &  J022317.32+381509.7   &  A1IV     & $254   \pm 35$    & $0.205 \pm 0.035$ \\
	AWI0005w7o  & 20994 &  J032429.84+341709.9   &  A8V      & $245   \pm 60$    & $0.174 \pm 0.030$ \\
	AWI0005yk3  & 19257 &  J030651.95+303136.8   &  F0V      & $79    \pm 4$     & $0.718 \pm 0.031$ \\
\enddata
\label{tab:w3}
\end{deluxetable}

\subsection{Binary Companions and Variability}

Five of the stars with newly reported excesses have known companions or secondaries.  None of these is apparent in the WISE images or likely to be cool enough to create the observed WISE infrared excesses. 

\begin{description}

\item[HD 7406] The separation of this binary is $\sim 1\arcmin$ \citep{1987A&AS...67..303S}, wide enough that the companion probably does not contaminate the WISE photometry.

\item[HD 74389]  HD 74389 has several nearby companion stars which could conceivably contaminate the SED. This star has a white dwarf companion discovered by \citet{1990PASP..102..440S}, type DA1.3, V=14.62 located $20.11\arcsec$ (2230 AU projected separation) to the west \citep{2013MNRAS.435.2077H}, which is readily visible in the DSS Blue, Red and IR images.  \citet{2013MNRAS.435.2077H} also report a nearby M dwarf companion to the west and a nearby sdB companion to the north.  
Indeed, the same DSS Blue, Red and IR images bands also show a second background object $\sim 13\arcsec$ to the north of the star, probably the sdB.

However, none of these objects is visible in the WISE or even the 2MASS images, which look like unsaturated clean point sources, probing a resolution at K band of $<4\arcsec$. Moreover, we only see a significant excess at W4, but not at W3.  The W3-W4 color of an M dwarf or sdB photosphere is $\sim$0 since the SEDs of these objects are in the Rayleigh Jeans limit beyond 10$\micron$, so the lack of a resolved companion at K band rules out a significant contribution to the W4 excess from the companion's photospheric emission.  An M dwarf or sdB unresolved by 2MASS might still be cause for concern if it were itself dusty.  But since the A star is the most luminous object in the system, it seems most likely that we are observing reprocessed light from dust around the A star.

See below for further discussion of this interesting object.

\item[HD 23873] is a member of the Pleiades, and listed as a binary in the Hipparcos Input Catalog \citep{1993BICDS..43....5T}. The separation of the binary is 11000 AU according to \citet{2001A&A...368..873M}, corresponding to roughly $1.5\arcmin$. 

\item[HD 207888] was identified as a visual binary by \citet{1988A&AS...75..515S}. The separation is wide enough ($\sim 1\arcmin$) that the companion probably does not comtaminate the WISE photometry.

\item[HD 22128] is a double-lined spectroscopic binary system with semimajor axis roughly 9 Solar radii. We found a spectral type of F0IV for the combined light, but in fact, the two components are A stars; see \citet{2013MNRAS.433.3336F} for a detailed spectral analaysis.  This star was first identified as a debris disk host star by \citet{2000PhDT........17S} based on ISO photometry (60--100$\mu$m).  We are the first to report excess emission from this source at WISE 4, excess at the 8-$\sigma$ level.   \citet{2007ApJ...663..365W} considered the object to be an ``anomalous system'' based on the high luminosity of its inferred disk given the star's age.  We note that the system appears slightly extended in 2MASS K, but not in other bands. No 18 $\mu$m flux measurement for this object appears in the Akari catalogue \citep{{2010A&A...514A...1I}}.  

\item[HD 224098]  This B8V star is listed as having faint optical secondary, type F0V, in \citet{1985A&AS...60..183L}.  No data on the separation of the secondary is provided.  

\end{description}

Two of our new disk candidates have known stellar variability of $<=0.04$ magnitudes in V band. This degree of variability is too small to affect our selection process.

\begin{description}

\item[HD 152308] is classified as an $\alpha2$ CVn variable star by \citet{2011MNRAS.414.2602D} and as type A0 Cr Eu according to \citet{2009A&A...498..961R}. The variability is 0.04 magnitudes in the Hipparcos system, with a period of 0.94 days.  Our classification yielded B9V.

\item[HD 218155]  This star varies over a range of $\sim0.03$ magnitudes in V band \citep{2006SASS...25...47W}.

\end{description}

\subsection{Be Stars}

Aside from normal main sequence stars, our program reveals four
new Be stars: HD 6612, HD 7406, HD 164137, and HD 218546. None of
these stars appears in the Be Star Spectra (BeSS) database
\citep{2011AJ....142..149N} or in the \citet{1999A&AS..134..255K} catalog of H$\alpha$
emission line stars, which cataloged stars with galactic latitude $|b|<10^{\circ}$.
All show prominent H$\alpha$ emission and the
strong upper level Balmer absorption lines characteristic of
classical Be stars \citep[e.g.,][]{2003PASP..115.1153P,2013A&ARv..21...69R}.
Two of these stars, HD 7406 and HD 218546, show significant W3 excess,
in addition to their W4 excess. The B1e star HD 7406 has additional strong He~I absorption lines;
we detect weak He~I absorption in the other stars.  Our high S/N
spectra reveal no trace of other permitted or forbidden emission
lines, confirming our classification of them as classical Be stars rather than
pre-main sequence Herbig Ae/Be stars \citep{2013A&ARv..21...69R,2013MNRAS.430.2169R}.

These stars illustrate the promise of Disk Detective for identifying new 
classical Be stars. Although Be stars have large infrared excesses due 
to free-free emission \citep[e.g.,][]{1974ApJ...191..675G,1994A&A...290..609D,2002AJ....124.3289B}, 
most have been identified from optical spectroscopic surveys \citep[e.g.,][]{1985ApJS...59..769S,2002AJ....124.3289B,2003PASP..115.1153P,
2013A&ARv..21...69R,2013MNRAS.430.2169R,2015MNRAS.446..274R}
optical photometric surveys \citep[e.g.,][]{2006ApJ...652..458W}
and IR spectroscopic surveys \citep[e.g.,][]{2015AJ....149....7C}.
Uniform selection based on infrared excess from WISE data
provides a flux-limited survey fairly independent of interstellar 
reddening. 

With only four Be stars, it is premature to analyze spectroscopic
properties and statistics for the WISE sample of classical Be stars.
However, it is worth noting that the frequency of late-type Be stars
(three B8--B9 stars and one B1 star) is somewhat larger than expected
from current samples where B4 and earlier stars are much more common
\citep[e.g.,][]{2002AJ....124.3289B,2013A&ARv..21...69R}. We plan a detailed analysis 
of a larger sample in a future paper (Bans et al., in prep).

\subsection{HD 74389: A Star With a Candidate Debris Disk and White Dwarf Companion}

One of our new candidate debris disks appears to orbit HD 74389 A, an A0V star with a white dwarf companion discovered by \citet{1990PASP..102..440S}. It also has a possible M dwarf companion and sdB companion reported by \citet{2013MNRAS.435.2077H}. There are three known planetary systems with white dwarfs as distant companions:
Gl 86 \citep{2000A&A...354...99Q, 2005MNRAS.361L..15M},
HD 27442 \citep{2001ApJ...555..410B, 2006A&A...456.1165C, 2006ApJ...646..523R, 2007A&A...469..755M},
and HD 147513 \citep{1997ApJ...476L..89P, 2004A&A...415..391M}.
However, the HD 74389 system appears to contain the first debris disk around a star with a white dwarf companion.

The white dwarf companion, HD 74389 B, has V mag 14.62, and is located 20.11 arcsec to the west \citep{2013MNRAS.435.2077H}.  It is readily visible in the DSS Blue, Red and IR images.  These bands also show a background object $\sim 13 \arcsec$ to the north of the star, possibly the M dwarf companion reported by \citet{2013MNRAS.435.2077H}.

It is interesting to ponder the origin of the disk around HD 74389 A and how the post-main sequence evolution of HD 74389 B may have affected it.  Though the white dwarf may presently have a projected separation of 2230 AU, it could have been 2-4$\times$ closer in when it was on the Main Sequence, thanks to stellar mass loss.  The disk may have merely survived the evolution of the higher-mass star, or it may represent a signature of dynamical changes to the system, like planet exchange \citep[e.g.,][]{2012ApJ...753...91K}.   It could even have been built from mass loss by the companion, via the process described by \citet{2013ApJ...764..169P}.

\section{DISCUSSION}
\label{sec:conclusions}

We have outlined and demonstrated a novel process for identifying new candidate circumstellar disks in the WISE survey data.  This paper reports only results from the first 10\% of the search, so it might be premature to try to derive any statistically meaningful inferences about the population of debris disk from this limited sample.  But our list of 37 new, well-vetted disk candidates demonstrates the utility of crowdsourcing analysis of WISE images.  One of our disk candidate systems appears to contain the first debris disk discovered around a star with a white dwarf companion: HD 74389.
We also report four newly discovered classical Be stars (HD 6612, HD 7406, HD 164137, and HD 218546) and a new
detection of 22 $\mu$m excess around previously known debris disk host star HD 22128.

We decided to only publish in this paper candidates that are in the Hipparcos catalog. Since the \citet{2013ApJS..208...29W} cross-correlated the WISE archive with the Hipparcos catalog, they
could conceivably have identified all of the candidates that we are announcing.
However, \citet{2013ApJS..208...29W} used a [2MASS] - [W4] color criterion, as opposed to our [W1] - [W4] color criterion, and used the WISE AllSky data rather than the WISE ALLWISE data. Yet all of the candidates presented in this paper
would have also been selected by the \citet{2013ApJS..208...29W} color criterion.
More importantly, while we examined candidates by eye to discard objects potentially contaminated by nearby stars
and galaxies, \citet{2013ApJS..208...29W} used a statistical likelihood-ratio (LR) technique to accomplish this goal. Perhaps their statistical technique was more conservative than our more labor intensive approach, leaving these candidates unidentified. 

We have several further improvements to Disk Detective project underway, which we will describe in upcoming papers, including: 
\begin{itemize}
\item New ways to retire the sources after fewer classifications.  
\item Spectroscopy of Southern hemisphere DDOIs via the CASLEO telescope in Argentina
\item Imaging follow-up of DDOIs with the Robo-AO \citep{2014ApJ...790L...8B} instrument at the Palomar Observatory 60-inch telescope to check for background contaminants located closer to the star than DSS can probe.
\end{itemize}
With its high sensitivity and angular resolution in the mid-infrared, we expect that the James Webb Space Telescope (JWST) will be an important tool for following up disks discovered via Disk Detective. So we aim to have the project mostly completed by the time JWST launches in the fall of 2018.

\acknowledgements

We acknowledge support from grant 14-ADAP14-0161 from the NASA Astrophysics Data Analysis Program. Marc Kuchner acknowledges funding from the NASA Astrobiology Program via the Goddard Center for Astrobiology.

Development of the Disk Detectives site was supported by a grant from the Alfred P. Sloan foundation, and the Zooniverse platform is supported by a Google Global Impact award.

WISE is a joint project of the University of California, Los Angeles, and the Jet Propulsion Laboratory (JPL)/California Institute of Technology (Caltech), funded by NASA. 2MASS is a joint project of the University of Massachusetts and the Infrared Processing and Analysis Center (IPAC) at Caltech, funded by NASA and the NSF. This paper uses data products produced by the OIR Telescope Data Center, supported by the Smithsonian Astrophysical Observatory.

The Digitized Sky Survey was produced at the Space Telescope Science Institute under U.S. Government grant NAG W-2166. The images of these surveys are based on photographic data obtained using the Oschin Schmidt Telescope on Palomar Mountain and the UK Schmidt Telescope. The plates were processed into the present compressed digital form with the permission of these institutions. This work has made use of the BeSS database, operated at LESIA, 
Observatoire de Meudon, France: http://basebe.obspm.fr

Funding for the SDSS and SDSS-II has been provided by the Alfred P. Sloan Foundation, the Participating Institutions, the National Science Foundation, the U.S. Department of Energy, the National Aeronautics and Space Administration, the Japanese Monbukagakusho, the Max Planck Society, and the Higher Education Funding Council for England. The SDSS Web Site is http://www.sdss.org/.

The SDSS is managed by the Astrophysical Research Consortium for the Participating Institutions. The Participating Institutions are the American Museum of Natural History, Astrophysical Institute Potsdam, University of Basel, University of Cambridge, Case Western Reserve University, University of Chicago, Drexel University, Fermilab, the Institute for Advanced Study, the Japan Participation Group, Johns Hopkins University, the Joint Institute for Nuclear Astrophysics, the Kavli Institute for Particle Astrophysics and Cosmology, the Korean Scientist Group, the Chinese Academy of Sciences (LAMOST), Los Alamos National Laboratory, the Max-Planck-Institute for Astronomy (MPIA), the Max-Planck-Institute for Astrophysics (MPA), New Mexico State University, Ohio State University, University of Pittsburgh, University of Portsmouth, Princeton University, the United States Naval Observatory, and the University of Washington.

The data presented in this paper result from the efforts of the Disk Detective volunteers, without whom this work would not have been possible. The following volunteers helped classify the stars listed in this paper: \small \textit{13lueAngeL,
A.Brodersen,
abienvenu,
AlarmingAlarm,
AlbeeLou,
alinastart,
anowi,
anoxie,
Arlon,
artman40,
arvintan,
atmosferah,
bc2callhome,
bellapagano,
berlherm,
biggsjrex,
billweiler,
bmw.996,
BraehlerM,
brainell,
breckwilhite,
Bromista,
cazze74,
Chinabob,
chloecollins ,
christania10110,
clairwallis,
clayzer-ev,
Cosmic Jerk,
cstickmaker,
daisy1000,
Deke scott T,
dianaz,
diplomacy42,
djperkins$@$live.co.uk,
DNiergarth,
dooces,
Ercydive,
feder,
ferrertiago,
fireandice,
firewatermoonearth,
Fletrik,
ftpol,
GBav8r,
geckzilla,
Gez Quiruga,
giazira,
golfmadman,
Grubenm,
Haian,
hardan,
hearzadrsos,
HelmutU,
i-mac,
iponenubs,
IrvSet,
Ivan\textunderscore3,
j.altman@kpnplanet.n,
jacedjohnson,
jacobus.valk$@$polk.de,
janiceashdown,
JasonJason,
jellybelly123,
JessicaElizabeth,
jgreyes,
jmeyers314,
JoJeFree,
jorge96,
josa310,
justinsbuzz,
kafter,
kecsap,
keel,
KI4FYP,
kiarash,
kijkuit,
Killoch,
kmk,
lambrosliamis,
lehensuge,
ljhinton21,
Mack777,
manemag,
marca,
marchl,
marcin.s,
Maria Mar,
mattlou117,
MDLW,
michiharu,
miertje,
miltonbosch,
mlavall2$@$uwo.ca,
Mric116,
MylesAtkins,
N5bz,
namorris,
nipper10,
nirajsanghvi,
norbertf,
Noreal,
nunolanca,
ohio\textunderscore pugmom,
OMHans,
onetimegolfer,
orionsam1,
Pamela Foster,
PaulRo,
peterw143,
Petrusperes,
Pini2013,
pixelfixx,
planetari7,
plutoexpress,
ptrip2010,
QGR,
Raymond Hall,
RefugeZero,
Ridence,
RobhJ,
Rocketman93,
RoLeCa,
rt26556$@$wdmtigers.or,
russ jones,
ryangeho,
ryanstone87,
sandrisvi,
scottwferg,
sheilaandpeter,
Shigeru,
shocko61,
Shroomzz,
silviug,
Siver,
SoloSlayer9,
Star hunter 1,
starbase3,
steve bourne,
SUMO\textunderscore 2011,
SunJinx,
symaski62,
TED91,
Tigrincs,
Timothy Fitzgerald,
tom.luthe,
Troomander,
Trumanator,
Tsgt,
turtle0920,
Victory1,
Vinokurov,
Vonkohon,
voyager1682002,
Vulpi,
weric1,
WizardHowl,
Woomaster,
WXdestroyer,
xantipa,
Yiska,
Zealex}  

We also thank Christoph Baranec and Katharina Doll for providing useful comments on this paper.

\section{APPENDIX: Additional Notes on Individual Disk Candidates}
\label{sec:appendix}

\begin{description}

\item[HD 9985] All four WISE images appear slightly ($\approx 2 \arcsec$ ) offset from the DSS and 2MASS images to the SW, a common feature of slightly saturated images in AllWISE.

\item[HD 14893]  \citet{2014yCat....1.2023S} lists this star as a B9.5 V.  We find type A0V.

\item[HD 22614] Possible member of the Pleiades cluster \citep{2001ApJS..132..253W}.

\item[HD 173056]  \citet{1999A&AS..137..451G} found a spectral type of A1V; we find A0.5V.

\item[HD 213290] DSS Blue, Red and IR show an unresolved background object roughly 14 arcsec north of the star. 

\item[HD 25466] Our spectral classification (A1V) matches that of \citet{2001A&A...373..625P} for this star. 

\item[HD 134854]  The SDSS images of this star are highly distorted.  

\item[HD 201377] Using higher resolution spectra (R=42000), \citet{2001A&A...369.1048P} find that this star is type A3 with logg=3.93. (We find A7V.)

\item[HD 20994] \citet{2015AJ....150...95A} identified this star in the Perseus OB 2 association as an AGB star based on its K-WISE 4 color alone.   However, our spectral typing (A8V) shows that this object is more likely to be a debris disk.

\item[HD 19257] The WISE 3 and WISE 4 excesses of this star have been reported by \citet{2013MNRAS.433.2334K}. We are the first to report a luminosity class for it; we find a spectral type of F0V, which disagrees with SIMBAD but is consistent with \citet{2013MNRAS.433.2334K}.

\item[HD 87827]  All four WISE images appear slightly ($\approx 2 \arcsec$ ) offset from the DSS and 2MASS images to the SW, a common feature of slightly saturated images in AllWISE.

\item[HD 221853]   We find a spectral type F2V for this well-known debris disk host star, consistent with the age estimate of 100 Myr in \citet{2015ApJ...798...87M}.

\end{description}


\end{document}